\documentstyle[12pt,twoside]{article}
\input epsf

\title{\Large\bf DETERMINATION OF LEADING TWIST NON--SINGLET OPERATOR 
MATRIX ELEMENTS BY QCD SUM RULES}
\author 
{
\it \bf  G. G. Ross$^1$, N. Chamoun$^2$ \\
\small$^1$  Department of Theoretical Physics, University of Oxford,\\
\small 1-4 Keble Road, Oxford, OX1 3NP, United Kingdom. \\
\small $^2$ Department of Theoretical Physics, University of Oxford,\\
\small 1-4 Keble Road, Oxford, OX1 3NP, United Kingdom. \\
}		
\date{}

\begin{document}
\maketitle
\begin{abstract}
We use QCD sum rules to determine the twist-two non--singlet operator
matrix elements and fixed $x$ structure functions paying particular
regard to the estimate of the errors. Particularly for the matrix
element determination, we find large uncertainties due to radiative
and higher dimension contributions. We consider the
origin of these large corrections and comment on their consequences
for other operator matrix element determinations. 
\end{abstract}

\section{Introduction}
QCD sum rules provide one of the few ways we have quantitatively  to
determine strong interaction effects and are applied widely when
extracting information from data. For example QCD sum rule estimates
for higher twist matrix elements have been used when testing the
Bjorken and Ellis Jaffe sum rules
\cite{kogan1,bbk,ross,stein}. However to our knowledge there has not
been a systematic study of the validity and accuracy of these
techniques when applied to structure functions. Given the importance
of the subject we consider this to be a serious omission and the
object of this paper is to provide just such a study. In particular we
will use QCD sum rules to determine the leading twist-two non-singlet
operator matrix elements (OMEs) contributing to deep inelastic
scattering 
from nucleon targets. In addition the errors to be expected from the
analysis 
are carefully analysed. The results will then be compared with the
precise  measurements of the leading twist matrix elements that are
now  available. This will allow
us not only to compare the QCD prediction with experiment but also to
test  our error estimates.

The determination of nucleon structure functions via QCD sum rules has
been pioneered by Belyaev and Ioffe \cite{ioffe-dis,ioffe-dis2}and
subsequently developed further by Balitsky et al \cite{bbk}. Indeed
Belyaev and Ioffe have presented the predictions for the non-singlet
structure functions themselves. To date we know of no analysis
of the leading twist OMEs themselves and, given
the current interest in determining specific OMEs
in various processes \cite{ross,stein,braun2}, we consider it
important so to  do. As we shall see there are significant differences
to be  expected in the accuracy of determination of the matrix
elements  compared to the determination of the structure functions at
fixed x.  We will also estimate the errors that arise from a variety
of  sources. The first is the inherent error following from the
uncertainty  of parameters such as the quark and gluon
condensates. The second  is the error that arises from neglect of
higher order  terms in the expansion in powers of $1/p^2
$ where p is the momentum flowing 
through the nucleon source (``higher--dimension'' terms). The third is
the error introduced when
estimating the left-hand-side of the sum rule by the resonance
contribution. Finally we consider the error following from the
perturbative expansion in powers of the QCD coupling. Of these it is
the second and forth terms that we find give the dominant source of
error in the determination of the 
OMEs and they are irreducible in the sense that there
appears to be no choice for the Borel transform scale M that is
consistent with the resonant saturation requirement while leading to a
convergent perturbative expansion and a well behaved ($\frac{1}{p^2}$)
expansion. We discuss the origin of this
problem and find it largely comes from the behaviour of the structure
functions near $x=1$. For this reason the  situation is better for the
determination of the structure functions at fixed x as originally
suggested by Belyaev and Ioffe \cite{ioffe-dis,ioffe-dis2}. We
reanalyse the fixed-x predictions, again making careful error
estimates. We find that even for the intermediate values of x which
have some convergence in the perturbative expansion the errors are
large. Moreover there is still a question concerning the large
corrections near $x=1$ for these are related to the quark not hadron
kinematics. As reliable predictions require some smearing of the
predictions before comparison with experiment is justified, we expect
some re-introduction of the sensitivity to the large perturbative
corrections at large $x$. Finally we comment on the origin of the
large  corrections and argue that they are likely to be present in the
determination of other interesting quantities including the
determination  of higher twist OMEs.
\section{Analysis}
To illustrate the difference between the calculation of the OMEs and
the  quark distribution, let us start with
the QCD sum rules for the quark distributions obtained in
ref. \cite{ioffe-dis} :\\
\begin{eqnarray}
\label{start1}
xu_{v}(x,Q^{2}) + M^{2} A^{\bar{\nu}p}(x,Q^{2}) &=&
\frac{M^{6}}{2 \bar{\lambda}^{2}_{N}}e^{m^{2}/M^{2}} L^{-\frac{4}{9}} \left\{ 
4E_{2}\left( \frac{W^{2}}{M^{2}} \right)x(1-x)^{2}(1+8x) \right.\nonumber\\
&&+\frac{b}{M^{4}} \left(
-\frac{4}{27}\frac{1}{x}+\frac{7}{6}-\frac{19}{12}x+\frac{97}{108}x^{2}
\right)E_{0}\left( \frac{W^{2}}{M^{2}} \right) \nonumber \\
&&\left. +\frac{8}{9}\frac{\alpha_{s}}{\pi}\frac{a^{2}}{M^{6}}
\left[
\frac{46}{9}-\frac{38}{9}x-2x^{2}-\frac{2}{9}\frac{x}{1-x}(1+14x)
\right] \right\}
\end{eqnarray}

\begin{eqnarray}
\label{start2}
xd_{v}(x,Q^{2}) + M^{2} A^{\nu p}(x,Q^{2}) 
&=&\frac{M^{6}}{2 \bar{\lambda}^{2}_{N}}e^{m^{2}/M^{2}}
L^{-\frac{4}{9}} \left\{
4E_{2}\left( \frac{W^{2}}{M^{2}} \right)x(1-x)^{2}(1+2x) \right. \nonumber \\
&&+\frac{b}{M^{4}} \left(
-\frac{4}{27}\frac{1}{x}+\frac{7}{6}-\frac{11}{12}x-\frac{7}{54}x^{2}
\right)E_{0}\left( \frac{W^{2}}{M^{2}} \right) \nonumber \\
&&+\frac{16}{9}\frac{\alpha_{s}}{\pi}\frac{a^{2}}{M^{6}} \frac{1}{1-x} 
\left[ x(1+x^{2})\left(\ln\frac{Q^{2}}{M^{2}x^{2}}+C-1 \right)
\right. \nonumber \\
&&
\left. \left. -\frac{8}{9} + \frac{13}{9}x + \frac{247}{36}x^{2}-6x^{3} \right]
\right\}
\end{eqnarray}
where $m$ is the nucleon mass, C is the Euler constant,
\[ a = -(2\pi)^{2}\langle 0 | \bar{\psi} \psi | 0 \rangle \]
\[ b = (2\pi)^{2} \langle 0 |
\frac{{\alpha}_{s}}{\pi}{G^a}_{\mu\nu}{G^a}_{\mu  \nu} 
	| 0 \rangle  \] 
\( E_{0}(z) = 1 -e^{-z} \), \( E_{2}(z) =1 -e^{-z}(1+z+\frac{1}{2}z^2)
\),
$W$ is the continuum threshold while \( L=\ln(M/\Lambda)/\ln(\mu/\Lambda) \)
takes account of the anomalous dimension of
the currents.
We express the $M^2$ dependence of $\bar{\lambda}^{2}_{N}(M^2)$
using
 the mass sum rule \cite{mass}
\[\bar{\lambda}_{N}^2 e^{-\frac{m^2}{M^2}} = M^6 L^{\frac{4}{9}} E_{2}\left(
\frac{W^{2}}{M^{2}} \right)+\frac{4}{3}a^2 L^{\frac{4}{3}}. \]
In equations (\ref{start1}) and (\ref{start2}), the first term on the
LHS comes from the resonant saturation with a nucleon pole and the
second parametrises some of the effects of the non--resonant
background corresponding to the interference of the pole term with a
continuum contribution. The
RHS comes from an evaluation of the deep--inelastic scattering process
in perturbative QCD. Note that the calculation keeps only leading
powers in an expansion in ($\frac{1}{Q^2}$) and hence projects onto
the leading twist contribution. For this reason, when comparing with
experiment, the comparison in \cite{ioffe-dis} was made at high $Q^2$.
However, this
introduces an error due to the neglect of the anomalous dimension
terms generating scaling ``violations''. As we presently discuss,
these are most easily included when calculating the OMEs. \footnote{ In
the case of the structure functions, the Altarelli--Parisi equations
should be used to evolve the result obtained from the QCD sum rules
for the structure functions to high $Q^2$ before comparing with data.}

The predictions for the structure functions following from
(\ref{start1}) and (\ref{start2}) are obtained in the usual way by
matching the LHS and RHS at fixed $x$ and $Q^2$. We shall return to
these shortly. First, however, let us consider the predictions for the
twist--2 OMEs. They follow immediately from the equations
(\ref{start1}) and (\ref{start2}) simply by taking moments:
\begin{eqnarray}
\label{defmomentu}
A_n^u	= \int_{0}^{1}dxx^nu_{v}(x,Q^2)
\end{eqnarray}
\begin{eqnarray}
\label{defmomentd}
A_n^d	= \int_{0}^{1}dxx^nd_{v}(x,Q^2)
\end{eqnarray}
Integrating the structure functions to obtain the moments as above is
equivalent
to evaluating the OMEs directly by the standard QCD sum rules method, 
inserting operators between nucleon currents. The only subtlety is
that the $\ln(\frac{Q^2}{M^2})$ term in eq. (\ref{start2}) is
replaced by $\ln(\frac{\mu_0^2}{M^2})$ where $\mu^2_0$ is the scale at
which the operator is renormalised. This should be chosen in a manner
that avoids higher order corrections,  we take ($\mu^2_0\simeq 1
GeV^2$) since the OMEs are taken between nucleon states. 

The twist--2 OMEs are now determined: 

\begin{eqnarray}
\label{umoment}
A_n^u + \frac{M^2}{m^2} R^{\bar{\nu}p}_n 
&=&\frac{M^{6}}{2 \bar{\lambda}^{2}_{N}}e^{m^{2}/M^{2}}
L^{-\frac{4}{9}} \left\{ 
4E_{2}\left( \frac{W^{2}}{M^{2}} \right)
\left(\frac{1}{n+1}+\frac{6}{n+2}-\frac{15}{n+3}+\frac{8}{n+4}\right)
\right.\nonumber\\
&&+\frac{b}{M^{4}} \left(
-\frac{4}{27}\frac{1}{n-1} + \frac{7}{6}\frac{1}{n}
-\frac{19}{12}\frac{1}{n+1} + \frac{97}{108}\frac{1}{n+2}
\right)E_{0}\left( \frac{W^{2}}{M^{2}} \right) \nonumber \\
&&\left. +\frac{8}{9}\frac{\alpha_{s}}{\pi}\frac{a^{2}}{M^{6}}
\left[
\frac{46}{9}\frac{1}{n} -\frac{38}{9}\frac{1}{n+1} 
+ \frac{-2}{n+2} + \frac{2}{9}\sum_{k=1}^{n}\frac{1}{k} 
+ \frac{28}{9}\sum_{k=1}^{n+1}\frac{1}{k}
\right] \right\}
\end{eqnarray}

\begin{eqnarray}
\label{dmoment}
A_n^d + \frac{M^2}{m^2} R^{\nu p}_n
&=& \frac{M^{6}}{2 \bar{\lambda}^{2}_{N}}e^{m^{2}/M^{2}}
L^{-\frac{4}{9}} \left\{ 
4E_{2}\left( \frac{W^{2}}{M^{2}} \right)
\left(\frac{1}{n+1}+\frac{-3}{n+3}+\frac{2}{n+4}\right) \right. \nonumber\\
&&+\frac{b}{M^{4}} \left(
-\frac{4}{27}\frac{1}{n-1} + \frac{7}{6}\frac{1}{n}
-\frac{11}{12}\frac{1}{n+1} + \frac{-7}{54}\frac{1}{n+2}
\right)E_{0}\left( \frac{W^{2}}{M^{2}} \right) \nonumber \\
&&+ \frac{8}{3}\frac{a^2}{M^6} \nonumber\\
&&+\frac{-16}{9}\frac{\alpha_{s}}{\pi}\frac{a^{2}}{M^{6}}
\left[
\frac{-8}{9}\sum_{k=1}^{n-1}\frac{1}{k} 
+ \frac{13}{9}\sum_{k=1}^{n}\frac{1}{k}
+ \frac{247}{36}\sum_{k=1}^{n+1}\frac{1}{k}
+ -6 \sum_{k=1}^{n+2}\frac{1}{k} \right. \nonumber \\
&& \left.\left. 
+ \left( \ln\frac{\mu_0^2}{M^2}+C-1 \right) 
\left( \sum_{k=1}^{n}\frac{1}{k} + \sum_{k=1}^{n+2}\frac{1}{k} \right)
 -2 \left( \psi'(n+1)+\psi'(n+3) \right)
\right] \right\}
\end{eqnarray}
where $\psi(a)=\frac{\Gamma'(a)}{\Gamma{(a)}}$ and $\Gamma(a)$ is the
Gamma function.\\
Since we are using dimensional regularisation, the singularities at
$x=1$ in eqs. (\ref{start1}) and (\ref{start2}) are not present,
$\frac{f(x)} {1-x}$ being replaced by the ``+ prescription''
$\frac{f(x)} {(1-x)_{+}} \equiv \frac{f(x)-f(1)} {1-x}$. The term
 proportional to $a^{2}$ in equation (\ref{dmoment})
corresponds to the $\delta(1-x)$ piece which was omitted in the
structure function expression in eq. (\ref{start2}) applicable at
intermediate values of $x$. 

In the following analysis we are particularly concerned with the
errors involved in the sum rules determination. We shall use the
following parameter  values
together with the quoted error estimates: 

\begin{itemize}
\item $a = 0.55 \pm 0.20 GeV^3$  \cite{ioffe-dis,gasser},
\item $b = 0.45 \pm 0.10GeV^4$  \cite{ioffe-dis,voloshin,narison},
\item $\mu=0.5 GeV$ which implies $\alpha_{s}a^2=0.13 GeV^6$ for three
flavour theory as in
\cite{ioffe-dis},
\item $\mu_0^2=1GeV^2$,
\item $\Lambda = 125 \pm 25 MeV$  consistent with the range of values
used in QCD sum rules; in ref.
\cite{ioffe-dis} it was taken to be 100MeV and in ref. \cite{mass}
it was taken as 150MeV,
\item $m=1.00 \pm 0.15 GeV$  \cite{mass},
\item $W^2=2.3 GeV^2.$ 
\end{itemize}
The uncertainty in the continuum threshold $W^2$ is taken to be of
order $10\%$ consistent with \cite{ioffe-dis} where the analysis
was done for two
choices of $W^2$: 2.3 and 2.5
$GeV^2$. However, as we shall see, increasing the uncertainty to
$100\%$ will not affect our phenomenological discussion (c.f. 
\cite{ross} where varying $W^2$ between 0.8 and 5 hardly affected the
results).

\begin{figure}
  \epsfxsize=14.cm
  \epsfysize=10.cm
  \centerline{\epsfbox{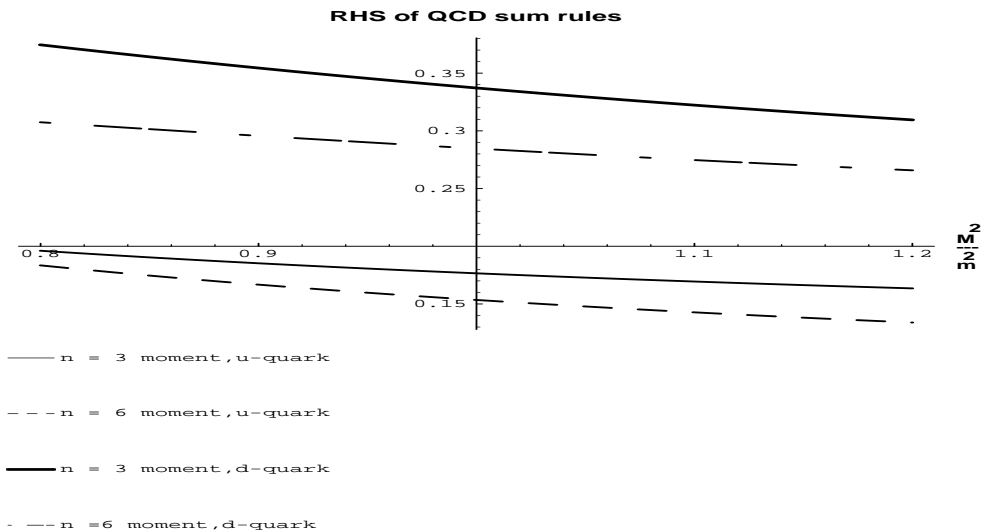 }}
  \caption{Plot of the RHS of the QCD sum rules for OMEs versus
$\frac{M^{2}}{m^2}$ showing the approximate linearity. }
  \label{fittingmoment}
\end{figure}

Let us first consider the QCD sum rules for the OMEs.
Figure ~\ref{fittingmoment} plots the RHS of the equations (\ref{umoment})
and (\ref{dmoment})
in the range $\frac{M^{2}}{m^{2}}\in[0.8,1.2]$ for the moments
3 and 6.  

According to the equations (\ref{umoment}) and (\ref{dmoment}), the
RHS  should be
linear in ($\frac{M^{2}}{m^2}$), and this may be seen to be
approximately true. To quantify this, we consider the differences
between a linear and a quadratic fit to the RHS. In the case of
u--quark (d--quark) a quadratic fit changes the constant term by
$46\%$, $23\%$ ($28\%$, $24\%$) for $n=6,\ 3$ respectively. This gives a
measure of the accuracy of the approximation of ignoring such terms in
the LHS of the QCD sum rules. In order to reduce this uncertainty, one
must show that the form of the continuum contributions are constrained
to prohibit quadratic terms of this magnitude. For clarity of
presentation, we shall not add this error to the errors found from the
uncertainties in determining the parameters used, but it should be
remembered that it must be included in the final error
estimates.

\begin{figure}
  \begin{tabular}{cc} 
   \epsfxsize=8.cm
   \epsfysize=8.cm
   \epsfbox{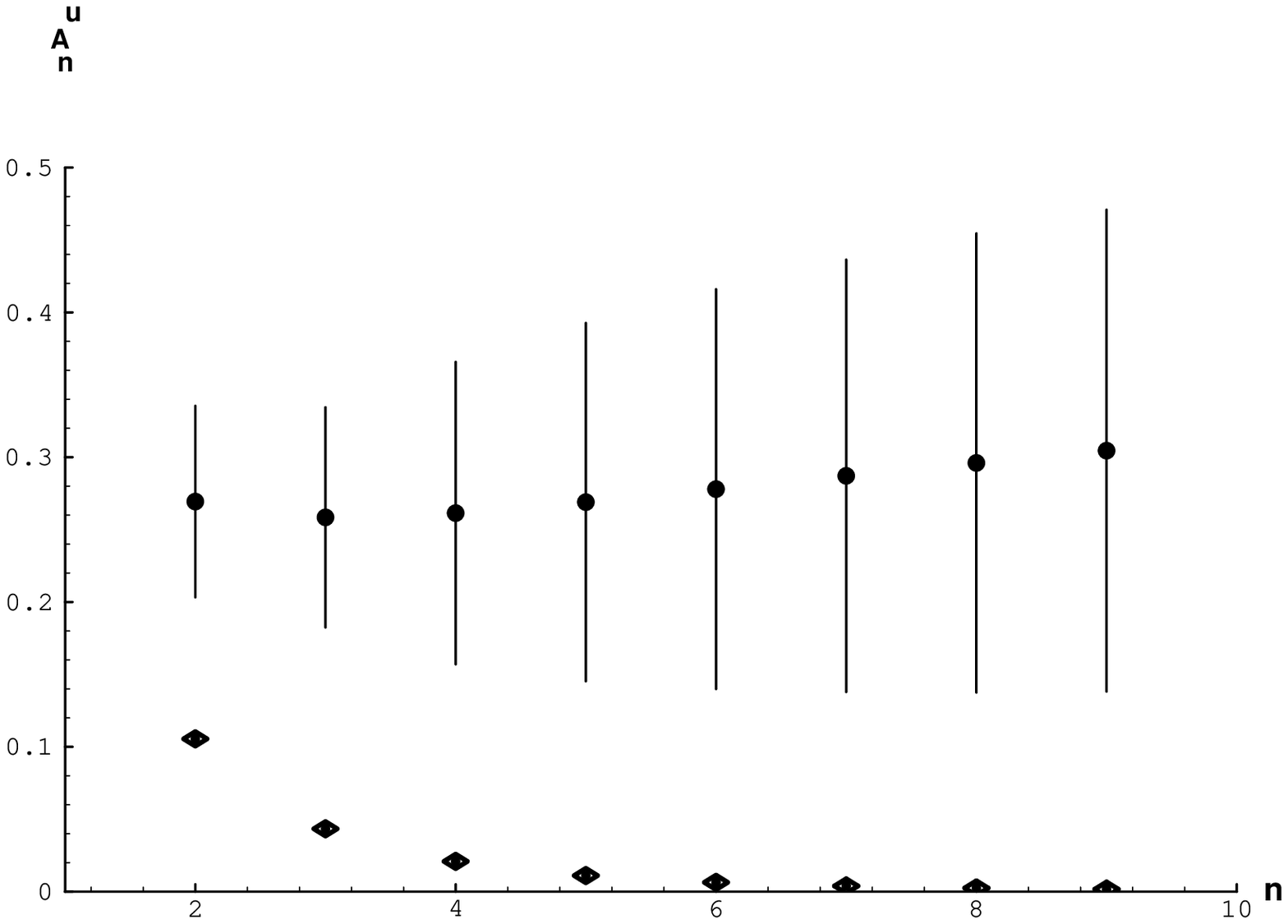 }&
   \epsfxsize=8.cm
   \epsfysize=8.cm
   \epsfbox{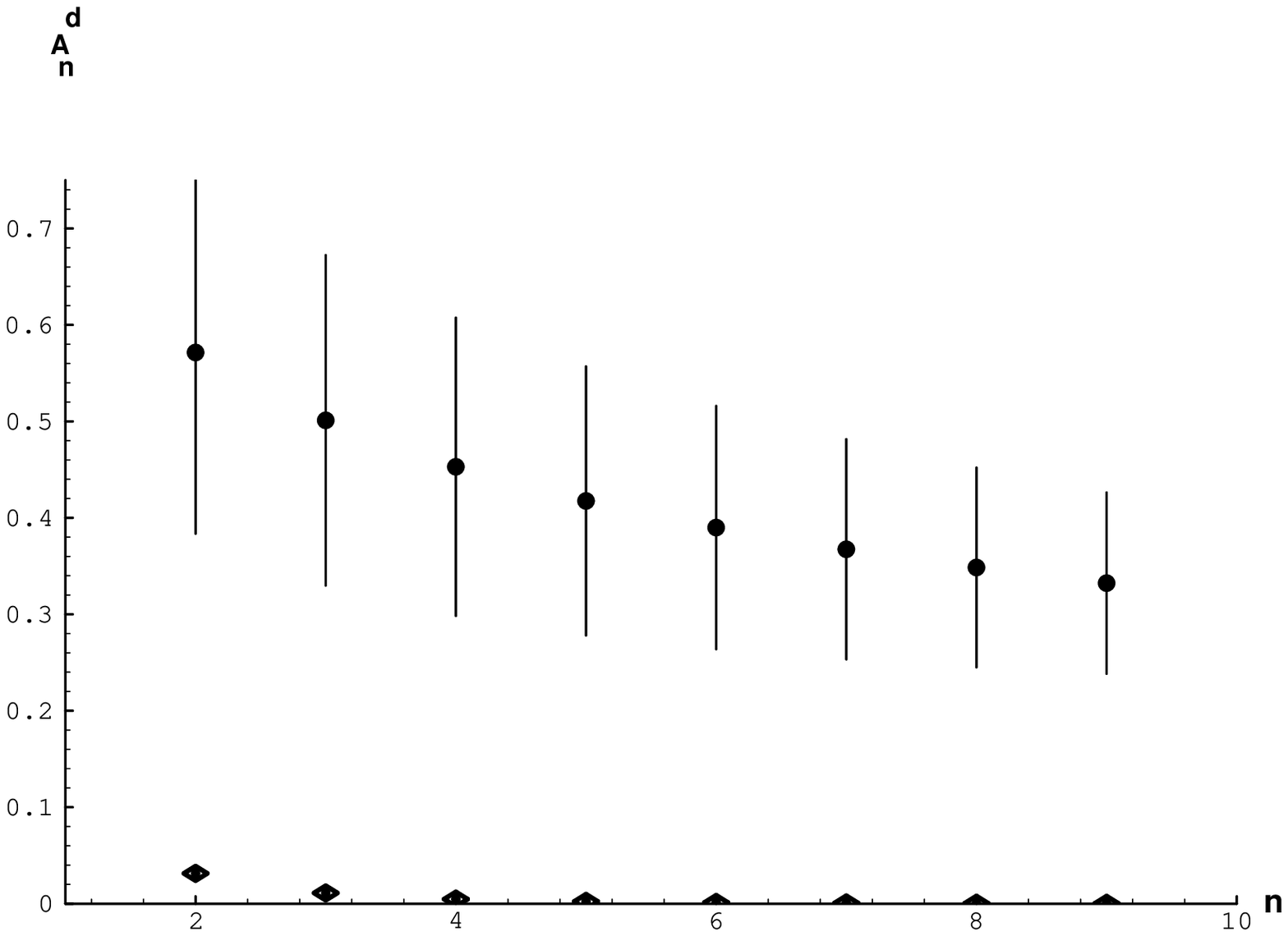 }\\
   (a)&(b)
  \end{tabular}    
    \caption{ Twist--2 non--singlet OMEs  following from QCD sum rules
together with the error estimates. Also shown are the experimental
determination of the OMEs (diamonds).}
  \label{figmoments}
\end{figure}

The second source of errors comes from the uncertainties in
parameters. By varying them over their allowed range and performing a
linear fit, we are able to determine the range of intercepts (the
OMEs) consistent within errors with the QCD sum rules. In Figs.
~\ref{figmoments}.a and
~\ref{figmoments}.b, we display the final results for the OMEs together
with their errors following from these
uncertainties. \footnote{Increasing the error in the continuum
threshold to $100\%$ makes a negligible difference to these results.} 
As discussed above, these are the twist--2 OMEs with normalisation scale
of order $1 GeV^2$. In order to compare with experiment, the moments
of the measured structure functions at high $Q^2$ are taken and fitted
to the contribution of all (leading and non--leading twist)
operators. At very high $Q^2$, the leading twist operators dominate
and so, using perturbative QCD to calculate the coefficient functions,
the leading-twist OMEs normalised at a scale of order $1 GeV^2$ are
readily obtained \cite{dick}.
   
The results obtained in Figs. ~\ref{figmoments}.a and ~\ref{figmoments}.b
are disappointing for the QCD sum rules method. The errors are very
large, yet the discrepancy with experiment is even greater. To explore
the origin of this discrepancy we consider the effect of the higher
dimension terms and the higher order 
perturbative expansion, by looking at the effect of adding
successive terms to the expansion, and comparing the result at each step.

\begin{figure}
 \begin{tabular}{cc} 
  \epsfxsize=8.cm
  \epsfysize=8.cm
  \epsfbox{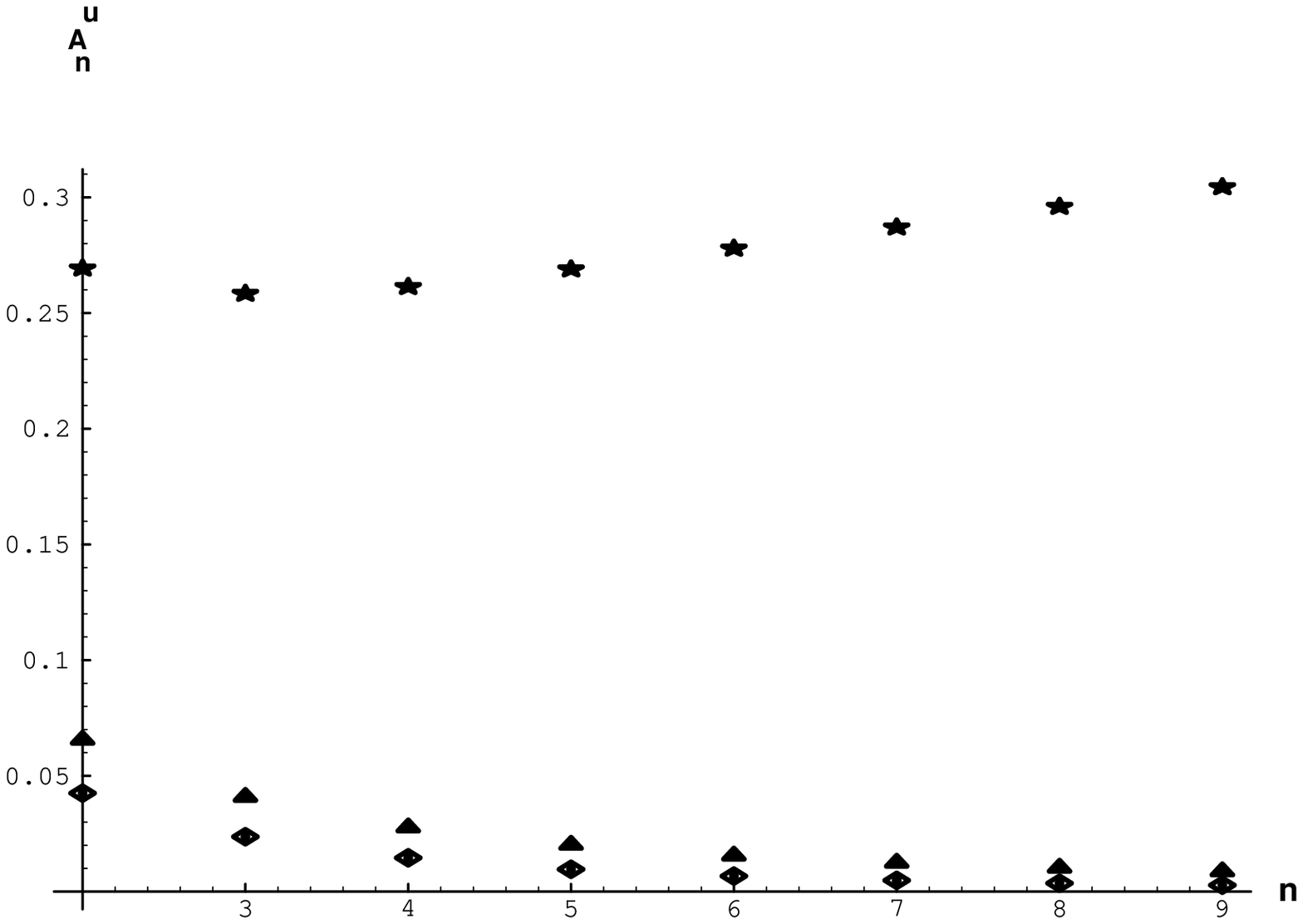}&
  \epsfxsize=8.cm
  \epsfysize=8.cm
  \epsfbox{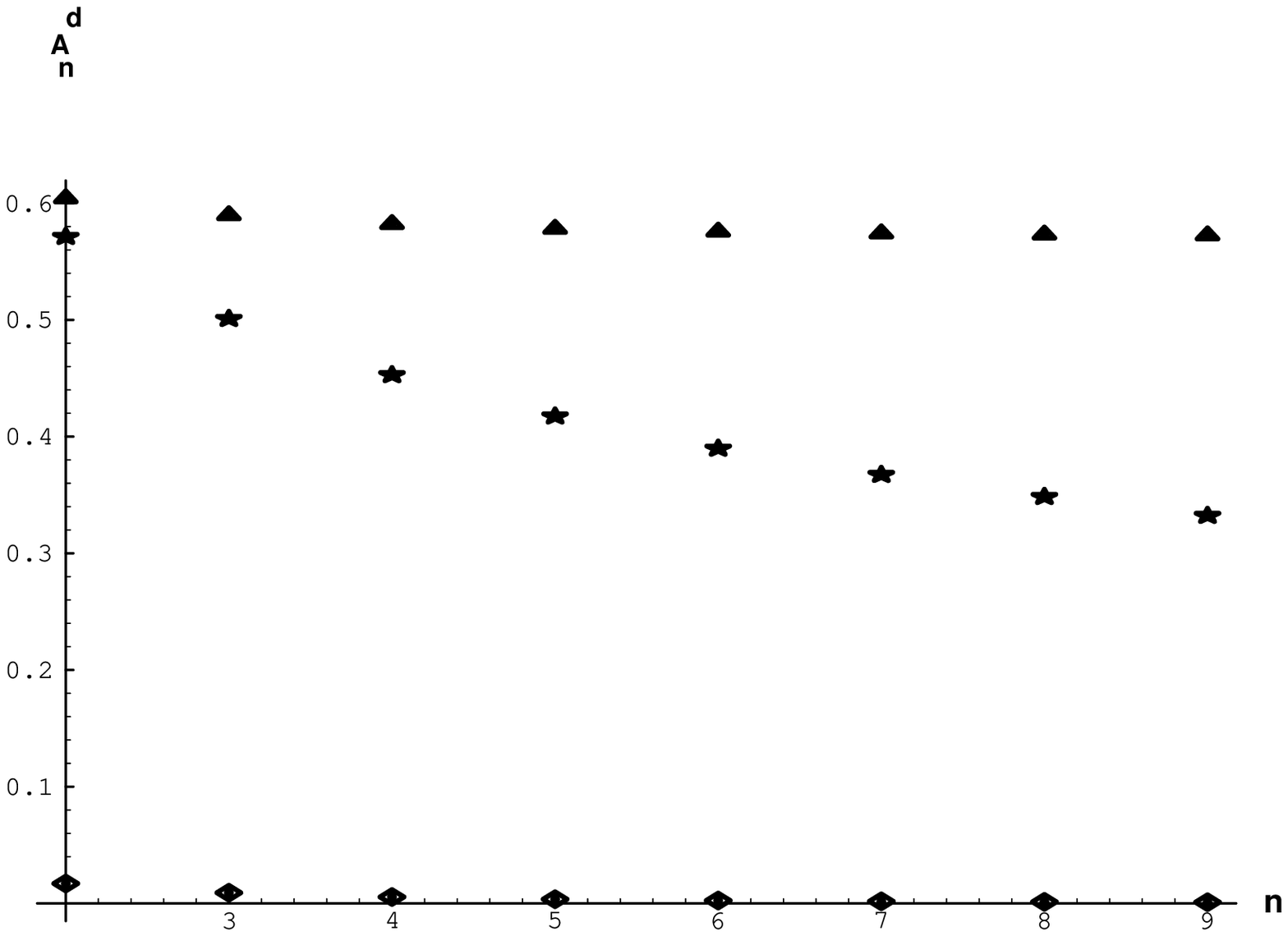 }\\
  (a)&(b)
 \end{tabular}
  \caption{Analysis of the importance of the various contributions to
the QCD sum rules. In (a)  we consider the
up--quark operators. The diamonds show the results neglecting the
gluon condensate and the $O(\alpha_s a^2)$ corrections. The triangles
show the results including the gluon condensate. The stars show the
complete result. (b)  shows the equivalent quantities for the
down--quark, although here the triangles include both the gluon and quark
condensates. Adding the gluon condensate alone changes the free quark
result imperceptibly).}
  \label{succmoment}
\end{figure}

The results are shown in Figs. ~\ref{succmoment}.a and
~\ref{succmoment}.b. From them we see that there is reasonable
agreement with experiment for the u--quark in the absence of the
$O(\alpha_s a^2)$ corrections. The latter contribution is peaked at
$x\simeq 1$ and affects all moments. For the d--quark, the agreement
with experiment is spoilt both by the $O(\alpha_s a^2)$ terms and by
the quark condensate which contributes a term $\propto \delta(1-x)$.
Again these terms contribute to all moments. The appearance of such
large perturbative and non--perturbative corrections to the free quark
model result signals a breakdown of the expansion assumed in deriving
the sum rules. For this reason, the error analysis presented is
inadequate, and one may understand why there is overall disagreement
between theory and experiment. It would be very interesting to
determine whether the alternate interpolating nucleon current used in
\cite{stein} improves the situation. 

\begin{figure}
  \epsfxsize=14.cm
  \epsfysize=10.cm
  \centerline{\epsfbox{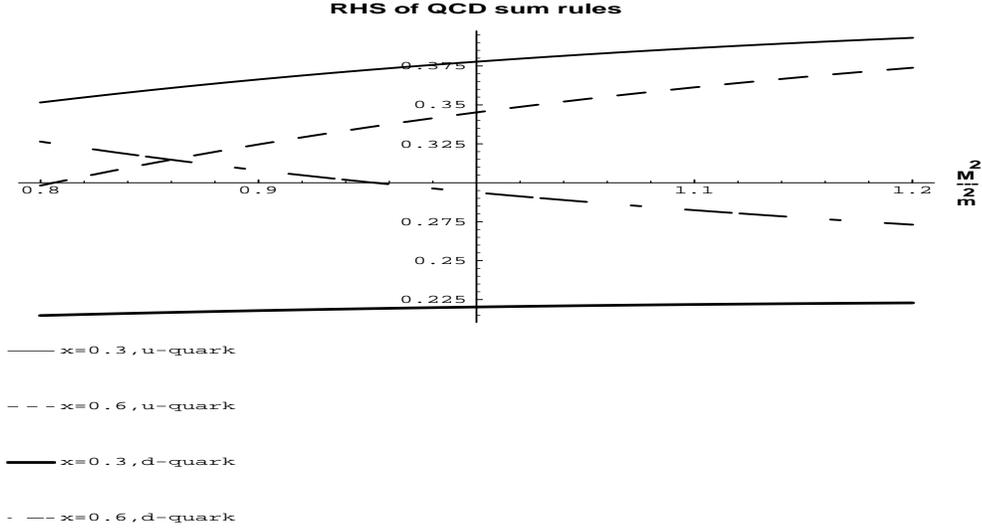 }}
  \caption{Plot of the RHS of the QCD sum rules for the structure
functions versus $\frac{M^{2}}{m^2}$ showing the approximate linearity.}
  \label{fittingstrfun}
\end{figure}

As the large corrections just discussed appear for large $x$, there is
a possibility that the predictions for structure functions are better
behaved and so we turn to a consideration of these effects
here. Figure  ~\ref{fittingstrfun} shows the RHS of the equations ~\ref{start1}
and ~\ref{start2}		
in the range $\frac{M^{2}}{m^{2}}\in[0.8,1.2]$ for the
values 0.3, 0.6 of $x$. From the RHS we may immediately determine the
quark distribution at each value of $x$.
Again using a linear and a quadratic fit, we may estimate the
contribution to the errors from the deviation from a straightline. In
the case of u--quark (d--quark) a quadratic fit changes the constant
term by $146\%$, $48\%$ ($34\%$, $20\%$) for the values $x=0.6, 0.3$
respectively.  This
gives uncertainties which again should be added to the
final errors. 

\begin{figure}
  \begin{tabular}{cc} 
   \epsfxsize=8.cm
   \epsfysize=8.cm
   \epsfbox{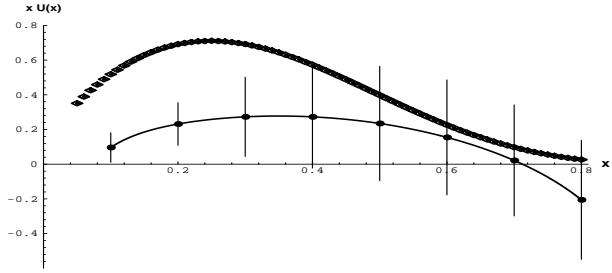 }&
   \epsfxsize=8.cm
   \epsfysize=8.cm
   \epsfbox{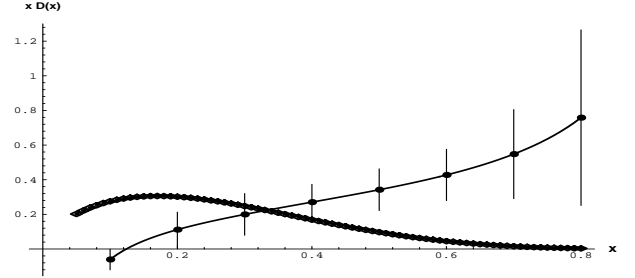 }\\
   (a)&(b)
  \end{tabular} 
  \caption{Quark structure functions following from QCD sum rules
together with the error estimates. Also shown are the experimental
determination of the structure functions (thick lines)}
  \label{strfun}
\end{figure}

Turning to the errors related to the uncertainties in the parameters,
we show in Figures ~\ref{strfun}.a and ~\ref{strfun}.b the
results for the up and down valence quarks. For comparison, the
observed twist--2 contribution to the structure functions is
shown--obtained by analysis of the large $Q^2$ data and continued
using QCD corrections to $Q^2=1 GeV^2$\cite{dick}.

\begin{figure}
 \begin{tabular}{cc} 
   \epsfxsize=8.cm
   \epsfysize=8.cm
   \epsfbox{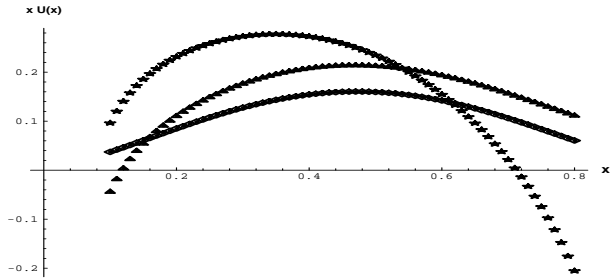 }&
   \epsfxsize=8.cm
   \epsfysize=8.cm
   \epsfbox{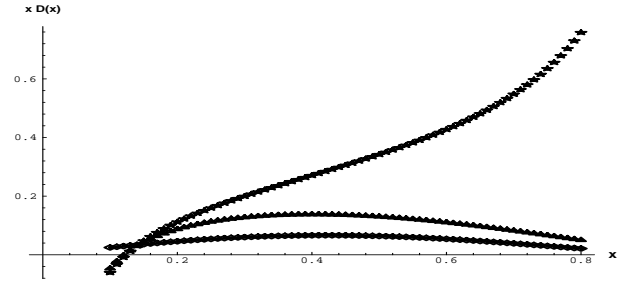 }\\
   (a)&(b)
  \end{tabular}
  \caption{Analysis of the importance of the various contributions to
the QCD sum rules. In (a) we consider the up-quark structure
function. The line of diamonds represents the free theory.
	The line of triangles represents the effect of adding the
gluon condensate $G^2$ and
 the line of stars  represents the effect of adding $G^2$ and
$\alpha_{s}a^2$. (b) shows the equivalent quantities for the down-quark.}
  \label{succstrfun}
\end{figure}

As before, we explore the effects of the higher dimension and higher
order terms. This is presented in
Figures ~\ref{succstrfun}.a and ~\ref{succstrfun}.b.
The region where the effect of adding the gluon condensate followed by the
$\alpha_s a^2$ perturbative term does not exceed $30\%$ corresponds to
$0.4 \leq x \leq 0.64$ for the u-quark and
$0.15 \leq x \leq 0.25$
for the d-quark. For the u--quark in the appropriate region, the
predictions  are consistent with
measurements within the errors. However, as one may see, the
theoretical errors are rather large.

\section{Summary and Conclusion}
The analysis shows that for the moments, in contrast to the structure
functions, there is no region of convergence for the OPE series. The
reason  is that the behaviour
at $x=1$ causes the power and$/$or perturbative expansion to break
down. For  the
u-quark, this was caused by the presence of the perturbative
term with component proportional to  $\frac{1}{1-x}$ regularized by
the ``$+$ prescription''. For the d-quark, the main problem is
caused by the term involving the 
$\delta(1-x)$ contribution.
It is these large terms
 at $x=1$ which cause the QCD sum rules analysis to break down. 

The origin of the $\frac{1}{(1-x)_{+}}$
terms which give the large $O(\alpha_s a^2)$ corrections is
particularly worrying for all QCD sum rules calculations of operator
moments. They arise from the exchange of a virtual gluon in graphs with a
single quark propagator. Such configurations occur for all OMEs
calculations. Given their importance in leading twist calculations, we
feel their magnitude should be computed to determine the reliability
of any OME calculation. In particular, the higher
twist calculations of \cite{stein} have not been done to this order
so some doubt needs to be cast on them. 

In the case of the structure functions, the effect of
these terms is apparently absent for values of $x \neq 1$. At least
for the u--quark, the predictions are in acceptable agreement with
experiment. However, the
question arises if the analysis is valid. In our opinion, this is
debatable, since the large corrections at $x=1$ arise from quark
kinematics and are not physical, so when
comparing to physical data, one can not assume their restriction to
$x=1$ as was done in \cite{ioffe-dis}. Indeed the concept
of duality tells us that
we should average over the cross section for production of quarks and
gluons in order to obtain the cross section of a
physical process, i.e. one should smear the singularities of quarks. This is
what the moments do, as  they average over the whole $x$ region of
integration $0 \leq x \leq 1$. While it may be possible to justify
averaging over a smaller range of $x$, any such averaging will
introduce sensitivity to the problematic behaviour at $x=1$.\\

{\bf Acknowledgement:} We are very grateful to R.G.Roberts for help in
this  analysis.

\end{document}